\documentclass{article}
\usepackage{graphicx}
\usepackage{epsfig}
\usepackage{epstopdf}
\usepackage{authblk}
\usepackage{subfigure}% subcaptions for subfigures
\usepackage{wrapfig}%   wrap figures/tables in text (i.e., Di Vinci style)
\usepackage[T1]{fontenc}
\usepackage[latin9]{inputenc}
\usepackage{units}
\usepackage{amsmath}
\usepackage{amssymb}
\usepackage{esint}
\usepackage{float}
\usepackage{longtable,tabularx}
\usepackage{color}
\begin{document}

%\title{Large eddy simulation study of the humidity variation in the shadow of a large wind farm in stable conditions}

\title{{\normalsize under consideration for publication in Wind Energy} \vspace {7mm} \\
 Large eddy simulation study of the humidity variation in the shadow of a large wind farm}

\author[1]{John Haywood*}

\author[1]{Adrian Sescu}

\author[2]{Kevin Adkins}

\affil[1]{Department of Aerospace Engineering, Mississippi State University, Mississippi, United States}

\affil[2]{Department of Aeronautical Science, Embry Riddle Aeronautical Engineering, Florida, United States}

\maketitle

%\corres{*\email{jsh478@cavs.msstate.edu}}

%\presentaddress{This is sample for present address text this is sample for present address text}

\abstract{Numerous studies have shown that wind turbine wakes within a large wind farm bring about changes to both the dynamics and thermodynamics of the atmospheric boundary layers (ABL). Previously, we investigated the relative humidity budget within a wind farm via field measurements in the near-wake region and large eddy simulations (LES). The effect of the compounding wakes within a large wind farm on the relative humidity was also investigated by LES. In this study, we investigate how the areas of relative humidity variation, that was observed in the near-wake, develop downstream in the shadow region of a large wind farm. To this end, LES of a wind farm consisting of 8x6 wind turbines with periodic boundary condition in the lateral direction (inferring an infinitely wide farm) interacting with a stable ABL is carried out. Two wind farm layouts, aligned and staggered, are considered in the analysis and the results from both configurations are compared to each other. It is observed that a decrease of relative humidity underneath the hub height and an increase above the hub height build up within the wind farm, and are maintained in the downstream of the farm for long distances. The staggered farm layout is more effective in keeping a more elongated region of low relative humidity underneath the hub, when compared to the aligned layout.}

%\keywords{Large Eddy Simulations, Wind Farm, Atmospheric Boundary Layer}

\section{Introduction}\label{sec1}

With the planned growth of renewable energy and the consequential deployment of a large number of wind turbines in many countries, research was focused on the climatic impact on the atmospheric boundary layer (ABL) \cite{Roy,Barrie}. Among other approaches, Large Eddy Simulation (LES) has been used extensively in the last decades to investigate the effect of large wind farm on the ABL (see, for example, Calaf et al. \cite{Calaf1}, Lu \& Port\'{e}-Agel \cite{Lu}, Yang et al. \cite{Yang}, VerHulst \& Meneveau \cite{VerHulst1}, Stevens et al. \cite{Stevens2}, Hayat et al. \cite{Hayat}). As the kinetic energy extracted from the upstream flow and from above the wind farm (the so-called 'entrainment', as was found in previous studies, such as Cal et al. \cite{Cal}, Meyers \& Meneveau \cite{Meyers}, Abkar \& Port\'{e}-Agel \cite{Abkar3}, VerHulst \& Meneveau \cite{VerHulst1,VerHulst2}) is used to generate power, downstream wake recovery from individual turbines or compounded wakes from multiple turbines is an important issue in the development of large wind turbine arrays. LES has shown that higher levels of upstream turbulence intensity aid in the recovery of wakes and moves the location of peak turbulence intensity and turbulent shear stress closer to the turbine \cite{Yu-Ting}. 

LES has been used to investigate the influence of atmospheric stability on entrainment, and the results pointed that the stable atmospheric scenarios lead to reduced entrainment \cite{Abkar1}. Alternatively, simulations point toward a weaker inversion strength or height \cite{Allaerts}, or increased positive buoyancy \cite{Abkar2}, increasing the entrainment rate and shortening the wake recovery. In an LES investigation of Calaf et al. \cite{Calaf1}, it was shown that, by neglecting stratification effects and specifically aiming at determining whether surface scalar fluxes change in the presence of wind turbines, there is an overall increase in scalar fluxes on the order of 10\%-15\% within a fully developed WTABL. Temperature fields within large wind farms in a stably or unstably stratified ABL have been also investigated using LES (see, for example, Calaf et al. \cite{Calaf}, Sescu \& Meneveau \cite{Sescu2}, Ali et al. \cite{Ali}). Generally, results agree with lower-resolution model studies that show how enhanced vertical mixing lowers the temperature above the rotor turbine top tip height and increases the temperature below the rotor turbine bottom tip height. Very few studies on the application of LES to predict the humidity budget in the ABL exist, however, and this study is an attempt to fill this gap (an example of a previous study is Adkins \& Sescu \cite{Adkins1}).
LES numerical experiments have also been utilized to explore the role of large-scale flow structures within the turbulent wake in entrainment (Cal et al. \cite{Cal}, Meyers \& Meneveau \cite{Meyers}, VerHulst \& Meneveau \cite{VerHulst1}). Such numerical experiments have also demonstrated how synthetic downward forcing of high velocity flow at upstream wind turbines can enhance kinetic energy entrainment and power extraction \cite{VerHulst2}.

Previous work using both experimental measurements and numerical simulation indicate that wakes generated inside large wind farms can substantially impact the exchanges of sensible heat and humidity within the ABL. More extensive investigations are necessary in order to understand and quantify these exchanges both within and downstream of a wind turbine array boundary layer (WTABL) as associated changes may impact human activities, especially in an agricultural context. For example, such impacts to crop production have been identified previously by Takle \cite{Takle}. Specific examples include: reduced summer moisture stress, reduced dew duration, spring soil drying, enhanced daytime photosynthesis, enhanced nighttime respiration, and increased moisture loss during drought (see, for example, Mortley et al. \cite{Mortley}, Pareek et al. \cite{Pareek}, Ford and Thorne \cite{Ford}, Tibbitts and Bottenberg \cite{Tibbitts1}, Tibbitts \cite{Tibbitts2}, or Grange and Hand \cite{Grange}).

In a previous study, Adkins and Sescu \cite{Adkins1,Adkins2} used experimental measurements to investigate the impact of a wind turbine on the relative humidity distribution in the near-wake region and numerical simulation to study relative humidity changes within a broader turbine array. Vertical, lateral and longitudinal observations allowed profiles of humidity in a stable ABL to be constructed. Vertical profiles with a decrease in relative humidity below the turbine hub height and an increase above were observed. Within the near-wake region, the relative humidity at the lower turbine tip height quickly decreased and slowly recovered with downstream distance. In the spanwise direction, at the lower turbine tip height, the greatest decreases in humidity were observed on the right-hand side of the wakeÕs centerline. This change was associated with the descending turbine blades on the right-hand side of the turbine disk and the wake interaction with the ground. In Adkins and Sescu \cite{Adkins1}, the agreement between the numerical LES results and the experimental measurements was found to be good although some of the comparisons were qualitative; however, we are confident that the comparisons can serve as a validation of the numerical tool that is being employed in this study, aimed at quantifying the effect of the wind farm shadow (the region that extends in the downstream of the wind farm for several kilometers) on the humidity budget in a stably stratified ABL.

The large-eddy simulation framework, the numerical algorithm employed to solve the equations, and the boundary conditions are outlined in the next section \ref{sec2}. In section \ref{sec3}, numerical results consisting of contour plots of various quantities or distribution of turbulent kinetic energy and relative humidity along different directions are presented and discussed. In the discussion, the focus is on how changes in relative humidity evolve in the shadow region of a wind farm, for both the aligned and staggered layouts.

\section{Large Eddy Simulation framework}\label{sec2}

As in Adkins \& Sescu \cite{Adkins2}, the LES filtered momentum conservation equations with the Boussinesq approximation, transport equations for potential temperature and specific humidity, as well as the continuity equation are employed,

\begin{eqnarray}\label{eq1}
\frac{\partial \tilde{u}_{i}}{\partial t}
+ \tilde{u}_{j}\frac{\partial \tilde{u}_{i}}{\partial x_{j}}
=
- \frac{\partial \tilde{p}^{*}}{\partial x_{i}}
- \frac{\partial \tilde{\tau}_{ij}}{\partial x_{j}}
+ \delta_{i3} g \frac{\tilde{\theta} - \langle \tilde{\theta} \rangle }{\theta_{0}}
+ f_c \epsilon_{ij3} (\tilde{u}_{j} - u_{gj})
+ F_{i} + F_{CPM}
\end{eqnarray}

\begin{eqnarray}\label{eq2}
\frac{\partial \tilde{\theta}}{\partial t}
+ \tilde{u}_{j}\frac{\partial \tilde{\theta}}{\partial x_{j}}
=
- \frac{\partial \pi_{j}}{\partial x_{i}} + F^{\theta}_{CPM}, 
\hspace{12mm}
\frac{\partial \tilde{q_s}}{\partial t}
+ \tilde{u}_{j}\frac{\partial \tilde{q_s}}{\partial x_{j}}
=
- \frac{\partial \pi_{j}^q}{\partial x_{i}}+ F^{q}_{CPM}, 
\hspace{12mm}
\frac{\partial \tilde{u}_{i}}{\partial x_{i}} = 0 
\end{eqnarray}
respectively, where the spatial filtering at scale $\tilde{\Delta}$ is represented by tilde, $\tilde{u}_{i}, i=1,2,3$, are the components of the velocity field corresponding to the axial $x_1$-direction, lateral $x_2$-direction, and vertical $x_3$-direction, respectively, $\tilde{\theta}$ is the resolved potential temperature, $\theta_{0}$ is the reference temperature, $\tilde{q_s}$ is the resolved specific humidity, the angle brackets represent a horizontal average, $g$ is the gravitational acceleration, $f_c$ is the Coriolis parameter, $\delta_{ij}$ is the Kronecker delta, $\epsilon_{ijk}$ is the alternating unit tensor, $\tilde{p}^{*}$ is the effective pressure divided by reference density, $F_{i}$ is a forcing term (here modeling the effect of the wind turbines), and the terms having $CPM$ at the subscript are active in a blending region, where the data from a precursor domain is transferred on to the main domain (more details below). The SGS stress, heat and humidity fluxes are given as $\tau_{ij} = \widetilde{u_i u_j} - \tilde{u}_i \tilde{u}_j$, $\pi_{j} = \widetilde{u_j \theta} - \tilde{u}_j \tilde{\theta}$, and $\pi_{j}^q = \widetilde{u_j q_s} - \tilde{u}_j \tilde{q_s}$. They are modeled using a Lagrangian scale-dependent model as developed by Bou-Zeid et al. \cite{Bou-Zeid}, and extended to scalar transport by Port\'{e}-Agel et al. \cite{Porte-Agel2}.

The numerical tool is a pseudo-spectral LES code that solves the filtered Navier-Stokes equations using a pseudo-spectral horizontal discretization and a centered finite difference method in the vertical direction (the grid is uniform in all directions) \cite{Calaf,Sescu}. Time marching is performed using a fully-explicit second-order accurate Adams-Bashforth scheme \cite{Butcher}. The continuity equation is enforced through the solution of the Poisson equation resulting from taking the divergence of the momentum equation. Periodic boundary conditions are imposed along the horizontal directions. The vertical gradients of velocity and the vertical velocity component vanish at the top boundary. The horizontal velocities at the first point away from the wall ($z = \Delta z/2$) are set through the velocity gradients in the vertical direction calculated using the Monin-Obukhov similarity theory, and the vertical velocity at the wall is set to zero. 

\begin{figure}[h]
 \begin{center}
  \includegraphics[width=13cm, clip=true]{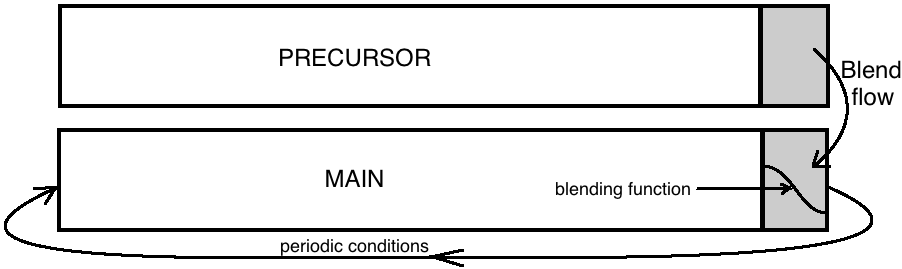}
 \end{center}
\caption{The procedure used to impose the inflow condition.}
\label{f1}
\end{figure}

A concurrent precursor simulation \cite{Stevens,Haywood} provides inflow boundary conditions that are introduced at the downstream boundary of the main domain in order to preserve the periodicity in the streamwise direction. It may sound counterintuitive to impose inflow conditions at the downstream boundary of the domain, but since periodic boundary conditions are imposed in the streamwise direction, whatever is imposed at the outflow boundary is automatically copied to the inflow boundary. The precursor and main flow domains considered here are identical, except the wind turbines rotors are added to the main domain. After each time step, a region of the flow data near the outflow boundary of the precursor domain is blended on to the flow data in a region located in proximity to the outflow boundary of the main domain. The blending region ensures that the flow data is smoothly transitioned from the precursor simulation to the main simulation. Assuming that the length of the blending region is $L_{blend}$ and ranges from $x = L_s$ to $x = L_x$, where $L_x$ is the length of entire domain, a generic variable (velocity, temperature, or relative humidity) in the blending region can be penalized using a source term in the governing equations ($F_{CPM}$, $F^{\theta}_{CPM}$, or $F^{q}_{CPM}$) to match the solution from the precursor domain; for example,
\begin{eqnarray}
F_{CPM} = w(x) \left[ \left( \tilde{u}_{i} \right)_{main} - \left( \tilde{u}_{i} \right)_{prec} \right]
\end{eqnarray}
where $\left( \tilde{u}_{i} \right)_{main}$ and $\left( \tilde{u}_{i} \right)_{prec}$ are flow variables from the main and precursor domains, respectively. The blending function used here is
\begin{eqnarray}
w(x) = \sigma 
  \begin{cases}
    \frac{1}{2}\left[1-cos\left(\pi \frac{x-L_s}{L_{pl}-L_s}\right)\right] &;\ L_s \leq x \leq L_{pl} \\
    1  & ;\ L_{pl}<x \leq L_x
  \end{cases}
\end{eqnarray}
where $L_{pl} = L_x - \frac{1}{4} L_{blend}$ (in the simulations included in this study, $L_{blend} = 0.05 L_x$, $L_s = L_x - L_{blend}$, and $L_{pl} = L_x - 0.25 L_{blend}$, where $L_x$ is the length of the flow domain in the streamwise direction) , and $\sigma$ is an amplitude, which is set equal to $0.05$ in this work. Figure \ref{f1} illustrates the procedure that is used to impose the inflow condition for the main simulation; the shape of the blending function that smoothly ramps the flow variables of the main simulation to the flow variables of the precursor simulation is also shown.

Due to the Coriolis effect, the direction of the wind changes with height in the ABL, subscribing to an Ekman spiral. This presents a challenge in trying to align the geostrophic velocity components to achieve the desired flow direction at hub height. For the simulations, an adjustment to the geostrophic wind direction is accomplished through manipulation of a Coriolis force type source term in the momentum equations (see equation (8) in Sescu \& Meneveau \cite{Sescu}) in order to achieve the desired hub height flow direction. Once the flow direction becomes normal to the rotor disk at the hub elevation, this term is deactivated to avoid unphysical behavior of the flow (more details about this procedure can be found in Sescu \& Meneveau \cite{Sescu}). 

An effective top layer of the ABL, isolated from the physically relevant flow within the ABL domain, is specified via a capping inversion created by a temperature gradient (the inversion strength is $0.01$ K/m). To this end, a source or sink of heat is introduced above the top of the ABL within the precursor simulation to enable the desired atmospheric stability (see figure 2 of Sescu \& Meneveau \cite{Sescu}). This is realized by including a source term in the scalar equation (see equation (4) in Sescu \& Meneveau \cite{Sescu}), where the amplitude is set via a PI controller with the input defined as the difference between the actual temperature and the desired temperature (see equation (5) and the following discussion in the same reference \cite{Sescu}).

An actuator disk method with rotation (ADM-R) similar to that employed by Wu and Port\'{e}-Agel \cite{Wu} is implemented here to model the effect of rotors on the ABL (more details about the implementation of the ADM-R can be found in Adkins and Sescu \cite{Adkins2}).

%\subsection{Example for second level head}

\section{Results and Discussion}\label{sec3}

%\subsection{Preliminaries}

\begin{figure}[H]
 \begin{center}
  \includegraphics[width=15cm, clip=true]{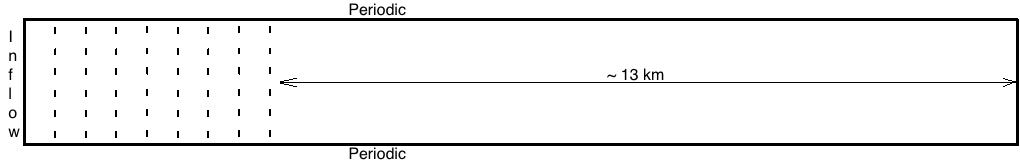}
 \end{center}
\caption{Wind farm layout with the shadow region.}
\label{f2}
\end{figure}

Two simulations were performed corresponding to an aligned and a staggered configuration, with the lateral spacing between two rotors of $416$ m, the longitudinal spacing of $680$ m, the rotor diameter set to $100$ m, the hub height equal to $80$ m, and a thrust coefficient of $0.6$. The sketch in figure \ref{f2} shows the aligned wind farm and the shadow region, the latter having a streamwise length on the order of $13$ km. A stable ABL with a thermal stratification of $2$ K and a geostrophic velocity of $8$ m/s, whose direction is maintained normal to the rotor disk at the hub height, are considered. A positive lapse rate of $0.5$ g/kg in the first 400 m was imposed for the specific humidity (closely resembling the humidity profile found from the experimental measurements of Adkins \& Sescu \cite{Adkins2}), where a constant potential temperature of 300 K and a constant specific humidity flux of $0.01$ g/kg m/s were imposed at the ground level. The distance between the inflow boundary and the first row of the wind farm is approximately $7$ rotor diameters (in the same order as the streamwise rotor spacing), which is sufficiently large to avoid unwanted induction. Simulations were performed within a domain having downstream, lateral, and vertical dimensions of $20$ km x $2.5$ km x $0.5$ km respectively, on a grid consisting of $512$ x $128$ x $96$ points. In the following, the relative humidity will be denoted by $q$, and its time average by $\bar{q}$.

%\subsection{Results}

\begin{figure}[H]
 \begin{center}
 a) \includegraphics[width=11cm, clip=true]{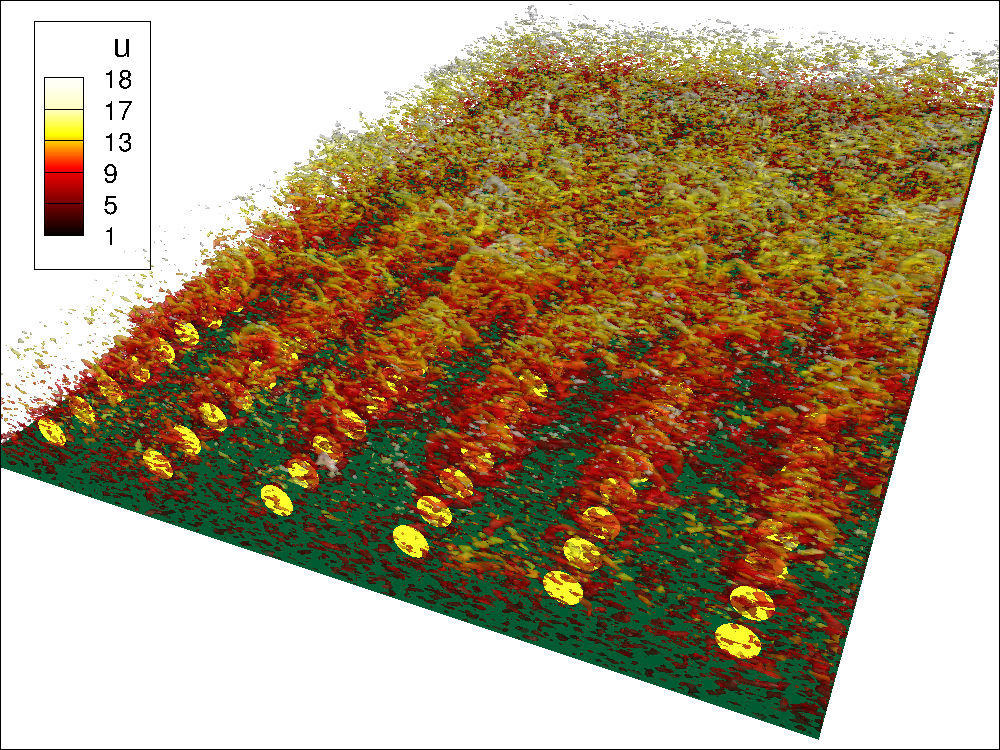} \\
 b) \includegraphics[width=11cm, clip=true]{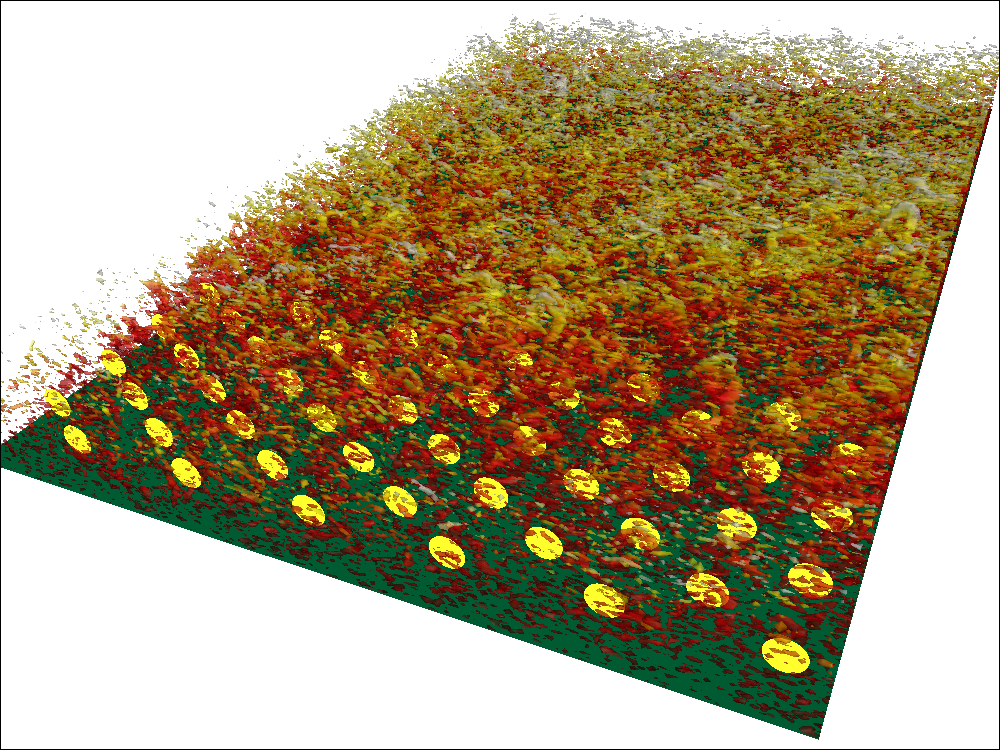}
 \end{center}
\caption{Qualitative representation of turbulence in the ABL via iso-surfaces of $Q$ criterion colored by the streamwise velocity component (the ground is shown in green, and wind turbine rotors are shown in yellow): a) aligned layout; b) staggered layout. The streamwise direction is scaled down by a factor of 10.}
\label{f3}
\end{figure}

\begin{figure}[H]
 \begin{center}
  \includegraphics[width=14cm, clip=true]{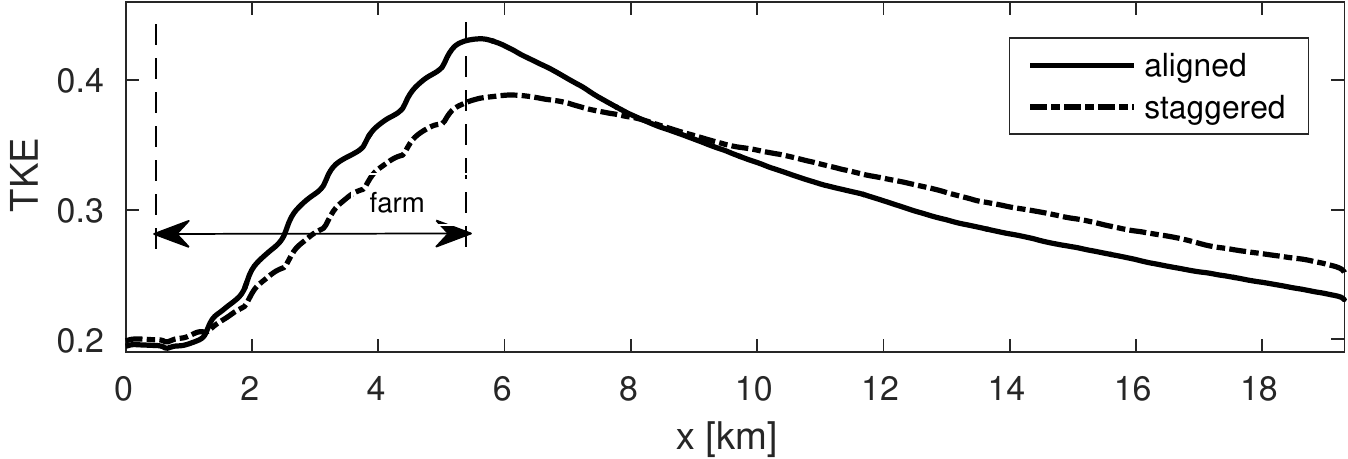}
 \end{center}
\caption{Turbulent kinetic energy (TKE) progression in the streamwise direction (TKE was integrated in the lateral and vertical directions). The blending region is not shown in this plot.}
\label{f4}
\end{figure}

Results in terms of iso-surfaces of Q-criterion, contour plots of mean relative humidity, and profiles of relative humidity in the longitudinal, lateral and vertical directions are presented and discussed next. The main objective is to analyze the effect of compounding wakes on near-surface relative humidity in the downstream of the wind farm (referred to as 'shadow region'). In figure \ref{f3}, we show iso-surface of the Q-criterion ($Q = 1/2[|\mathbf{\Omega}|^2 - |\mathbf{S}|^2]$, where $\mathbf{S} = 1/2[\nabla \mathbf{v}+(\nabla \mathbf{v})^T]$ is the rate-of-strain tensor, and $\mathbf{\Omega} = 1/2[\nabla \mathbf{v}-(\nabla \mathbf{v})^T]$ is the vorticity tensor) colored by the streamwise velocity component, where the ground is in green, and wind turbine rotors are in yellow (the streamwise direction is scaled down by a factor of 10). We decided to use Q-criterion as opposed to vorticity magnitude because it is one of the most effective approaches of vortex identification, and it better highlights the turbulence intensity in the wakes, as well as gives a qualitative representation of the level of mixing. The effect of compounding wakes is more prevalent for the aligned layout in figure \ref{f3}a than that corresponding to the staggered layout in figure \ref{f3}b. Turbulence in the upstream of the wind farm is characteristic of a stable ABL, with small flow structures convected by the base flow - the thermally stratified ABL tends to suppress large flow structures; the appearance of turbulence in the downstream, however, changes since the wind turbine rotors are efficient mixers, generating more intense flow structures, thus increasing the turbulence kinetic energy. 

The  turbulence kinetic energy (TKE) integrated in the lateral and vertical directions as

\begin{eqnarray}\label{eq5}
TKE(x) = \frac{1}{z_h(y_2 - y_1)} \int_{0}^{z_h}\int_{y1}^{y2} \left[ \overline{u'^2}(x,y,z) + \overline{v'^2}(x,y,z) + \overline{w'^2}(x,y,z) \right] dy dz,
\end{eqnarray}
where $y_1$ and $y_2$ are the coordinates of lateral boundaries, $z_h$ is the height of the domain, and bars represent time average of the fluctuating velocity components $u'$, $v'$ and $w'$, is plotted as a function of the streamwise direction in figure \ref{f4}; both curves corresponding to the two farm layouts were superposed to each other. The graph clearly indicates that the integrated TKE reaches a maximum at the end of the wind farm (actually, slightly downstream by roughly a streamwise spacing between two consecutive rotors), after which it experiences a slow decrease in the shadow region. The integrated TKE corresponding to the staggered layout (dashed line) increases at a smaller rate within the farm and decreases at a smaller rate in the shadow region, but reaches a higher level further in the downstream, suggesting a longer recovery region. The reason for which the wakes recover faster in case of the aligned layout is the higher turbulence intensity in the wakes and therefore the creation of a stronger vertical kinetic energy flux. This contrast between the aligned and staggered layouts has been identified in a number of previous studies (see, for example, Chamorro et al. \cite{Chamorro}, VerHulst \& Meneveau \cite{VerHulst1,VerHulst2}, or Stevens et al. \cite{Stevens3,Stevens3}). The higher peak in the TKE for the aligned layout is due to the effect of compounding wakes that enhance mixing and turbulence intensity more effectively, because the distance between two consecutive turbines is smaller when compared to the staggered layout.

\begin{figure}[H]
 \begin{center}
  \includegraphics[width=15cm, clip=true]{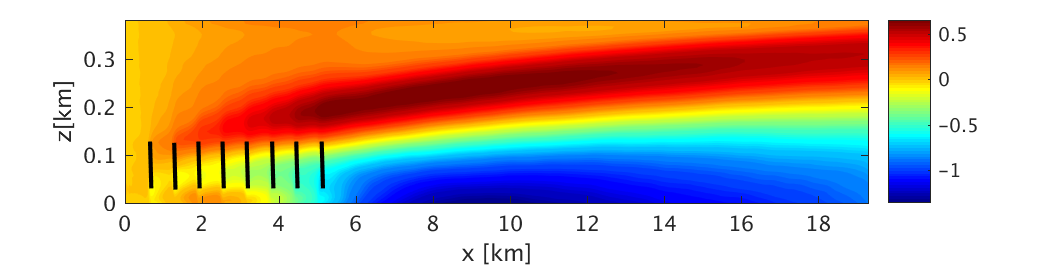} \\
  \includegraphics[width=15cm, clip=true]{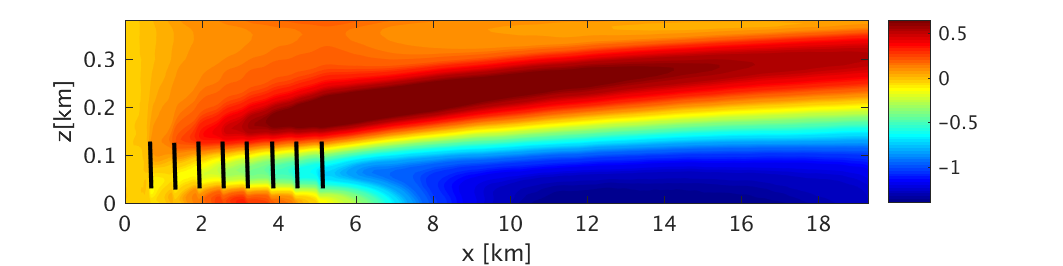}
 \end{center}
\caption{Mean relative humidity progression (calculated using equation (\ref{eq5}) as a percentage difference) contour plots: aligned layout (top); staggered layout (bottom). The streamwise direction is scaled down by a factor of 12, and the rotors are represented by 8 vertical lines).}
\label{f5}
\end{figure}

Next, the behavior of the relative humidity in the shadow region is studied qualitatively and quantitatively. First, contours of the time-averaged and spanwise-averaged relative humidity progression calculated as

\begin{eqnarray}\label{eq5}
\Delta q(x,z) = \frac{1}{T(y_2 - y_1)} \int_{0}^{T}\int_{y1}^{y2} \left[ q(x,y,z,t) - q_{up}(y,z,t) \right] dy dt,
\end{eqnarray}
where $q_{up}(y,z,t)$ is the humidity two diameters upstream of the first row of turbines, and $T$ being a sufficiently long time window (in the order of $1.5$ hours, which corresponds to approximately two flow through times),
 are plotted in figure \ref{f5} for both the aligned (top) and staggered (bottom) layouts (in this figure, the streamwise coordinate is scaled down by a factor of 12, and the rotors are represented by 8 vertical, black lines). Even with this time window, there where, however, some grid-to-grid oscillations in the reported results that have been eliminated via high-order filters, which do not compromise the validity of the results. In the shadow region, both parts of the figure strongly indicate that there is a reduction of the relative humidity in proximity to the ground, which extends up to the total height of the wind turbines ($\sim130$ m), and that there is an increase above the wind farm. The region of increase gradually moves upward with the streamwise direction reaching altitudes in the order of twice the total wind turbine height ($\sim300$ m). It is interesting to note that the decrease in the relative humidity in the vicinity of the ground continues in the downstream of the wind farm for almost another length of the farm (from $6$ km to $10$ km for the aligned layout and from $6$ km to $11$ km for the staggered layout). The recovery of the relative humidity under the hub (recovery meaning the convergence of the humidity to the upstream levels) seems to be faster for the aligned layout as indicated by the top panel of figure \ref{f5}. 

\begin{figure}[H]
 \begin{center}
  \includegraphics[width=12cm, clip=true]{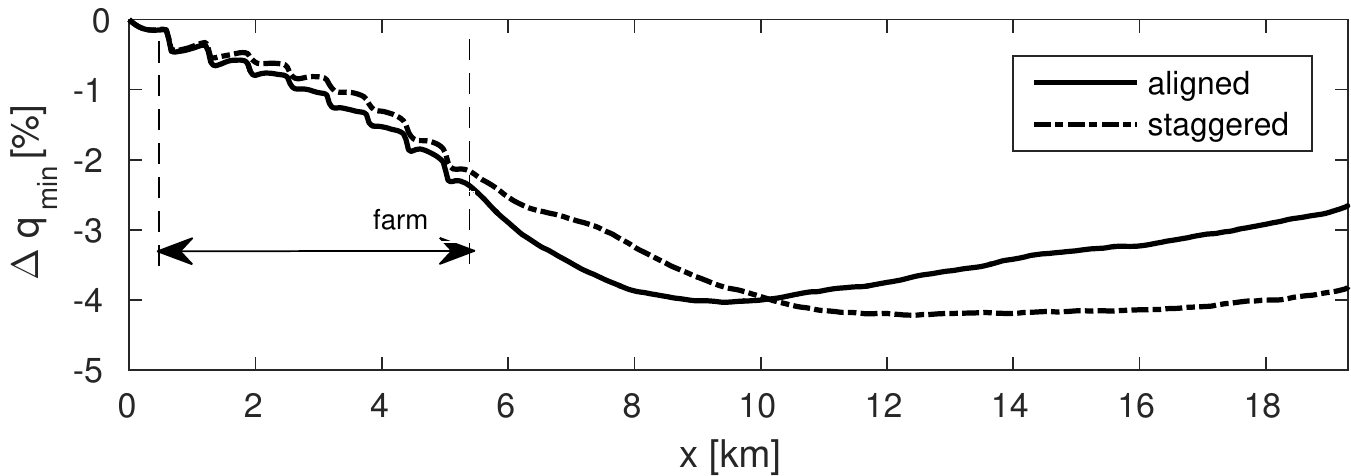} \\
  \includegraphics[width=12cm, clip=true]{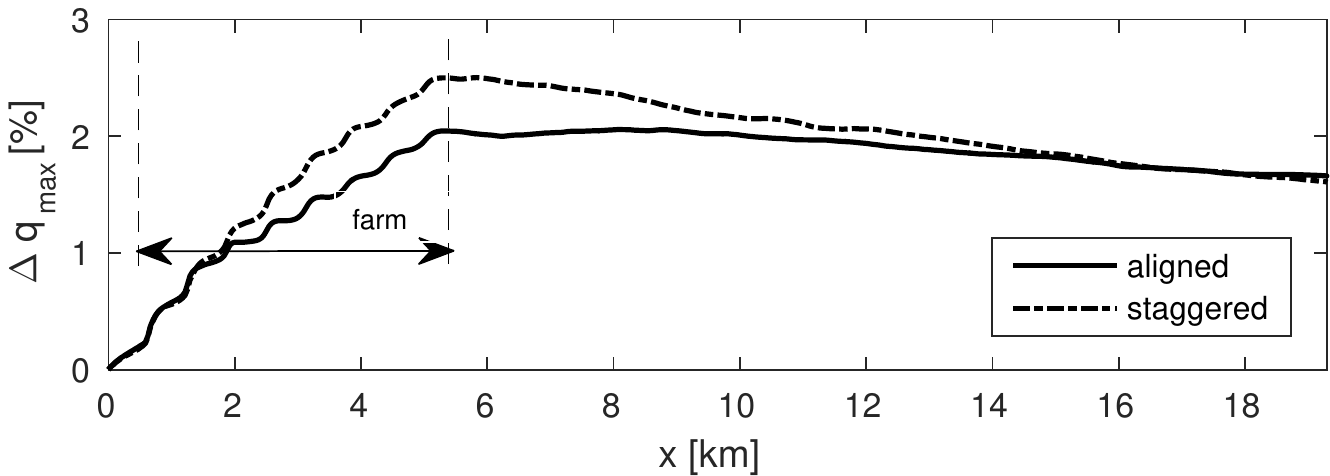}
 \end{center}
\caption{Mean relative humidity progression in the streamwise direction: decrease - under the hub (top); increase - above the hub (bottom). The blending region is not shown in this plot.}
\label{f6}
\end{figure}

The variation of the relative humidity in the streamwise direction evaluated as

\begin{eqnarray}\label{eq6}
\Delta q_{min}(x) = \frac{1}{y_2 - y_1} \int_{y1}^{y2} \min_{0<z<z_h} \left[ \bar{q}(x,y,z) - \bar{q}_{up}(y,z) \right] dy,
\end{eqnarray}

\begin{eqnarray}\label{eq7}
\Delta q_{max}(x) = \frac{1}{y_2 - y_1} \int_{y1}^{y2} \max_{0<z<z_h} \left[ \bar{q}(x,y,z) - \bar{q}_{up}(y,z) \right] dy,
\end{eqnarray}
where bar denotes the time average
$
\bar{q}(x,y,z) = 1/T \int_{0}^{T} q(x,y,z,t) dt
$,
 is quantitatively analyzed in figure \ref{f6}; the minimum values are representative of the humidity decrease under the hub and the maximum values are representative of the humidity increase above the hub. Apparently, the decrease in relative humidity under the hub (top panel of figure \ref{f6}) evolves differently in the streamwise direction for the two layouts: the humidity corresponding to the aligned layout drops at a higher rate between $x=6$ km and $x=8$ km than the rate corresponding to the staggered layout; then, it recovers again at a higher rate in the downstream, while the recovery of the staggered layout is much smaller. This can be correlated with our previous assertion (in the paragraph before figure \ref{f5}) about the behavior of the kinetic energy in the wake, which recovers faster for the aligned layout. Above the hub height (bottom panel in figure \ref{f6}) the behaviors of the relative humidity variations are not much different between the two farm layouts: both experience a steady decrease in the shadow region, though at a slightly smaller rate for the aligned layout. The results in figure \ref{f6} indicate that the observed trends in the wind farm, with respect to the variation of integrated relative humidity both below and above the hub, are similar as in the previous work of the authors (see figure 9 in Adkins \& Sescu), except a slight difference in the levels because the turbine spacings and the thrust coefficient are different.

\begin{figure}[H]
 \begin{center}
  \includegraphics[width=7.cm, clip=true]{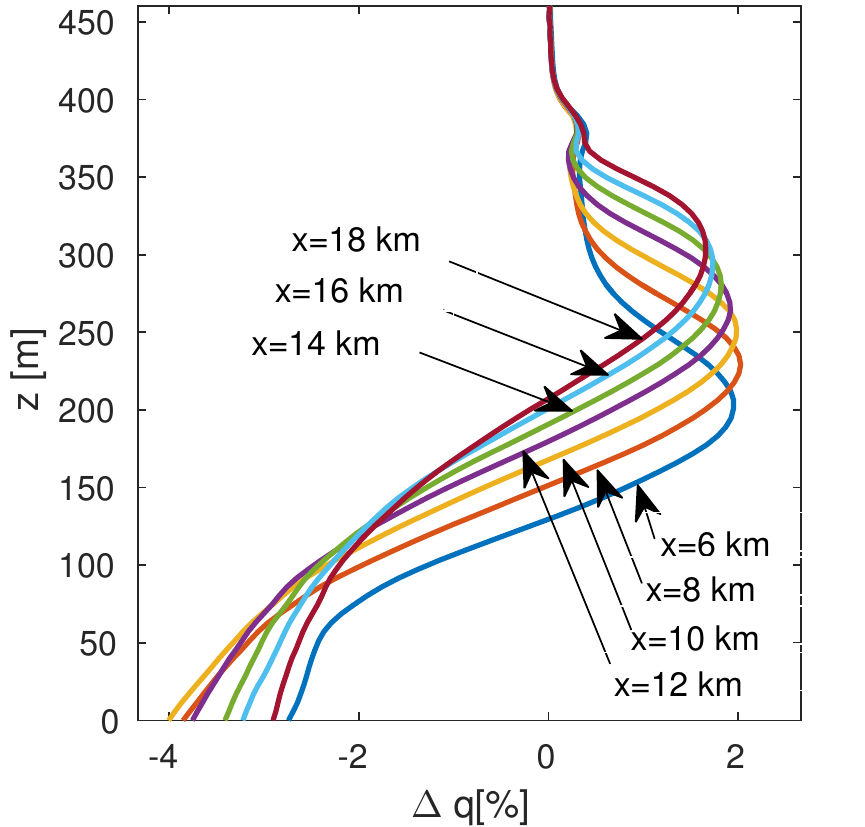}
  \includegraphics[width=7.cm, clip=true]{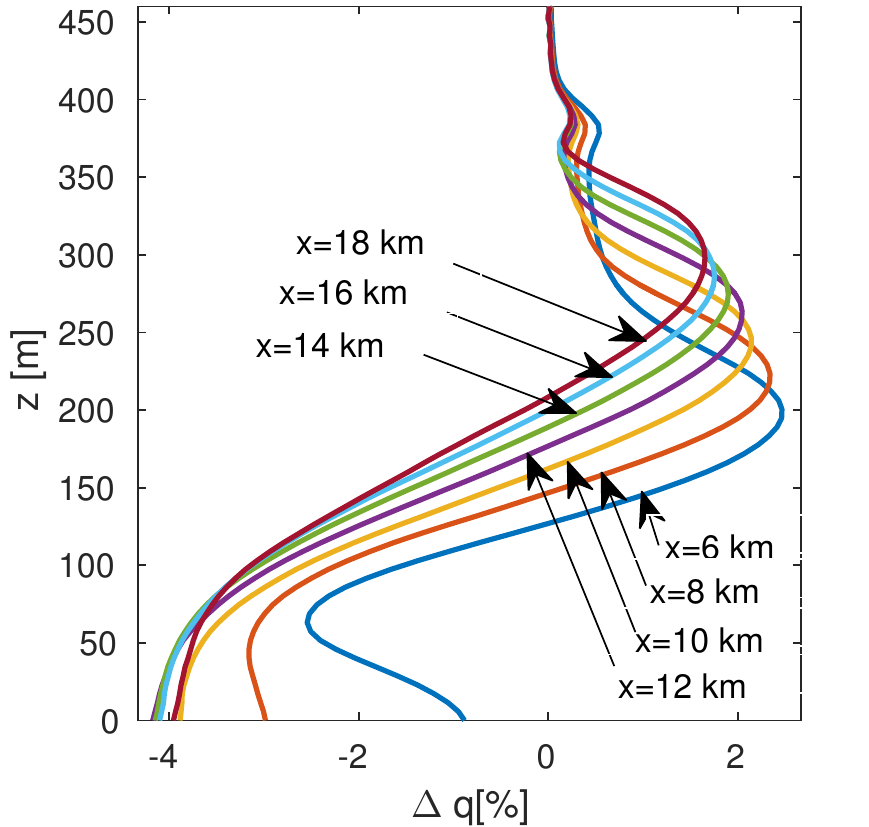}\\
  a)  \hspace{70mm} b)
 \end{center}
\caption{Vertical profiles of the mean relative humidity for different streamwise locations: a) aligned layout; b) staggered layout. $x=6$ km corresponds to the end of the wind farm.}
\label{f7}
\end{figure}

In figure \ref{f7}, we plot vertical profiles of the time-averaged and spanwise-averaged relative humidity variation (see equation \ref{eq5}) at selected streamwise locations (they are in fact a quantitative representation of the contour plots shown in figure \ref{f5}). Both panels in figure \ref{f7} show that the location of zero humidity variation (where the switching from negative to positive values occurs) gradually moves upward with the increase in the streamwise location, and that this location is well above the hub height: it varies between $130$ m at $x=6$ km and $200$ m at $x=18$ km. Both layouts bring the relative humidity variation to the same level at the ground, with a faster recovery for the aligned layout as previously concluded from the contour plots shown in figure \ref{f5}. Comparing the progression of the relative humidity in the upper layers, it seems that the staggered farm configuration yields a higher increase in the immediate proximity to the farm, but then it decreases at a faster rate in the downstream.
 
\begin{figure}[H]
 \begin{center}
  0.1 D \includegraphics[width=12cm, clip=true]{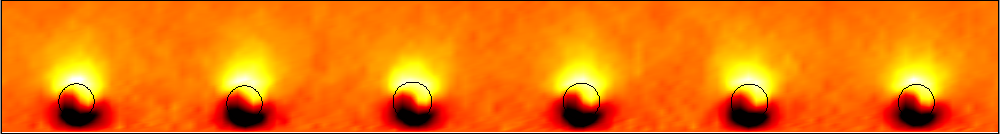}\\ \vspace{0.5mm}
  0.5 D \includegraphics[width=12cm, clip=true]{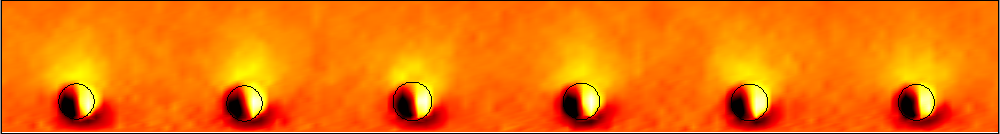}\\ \vspace{0.5mm}
  1.0 D \includegraphics[width=12cm, clip=true]{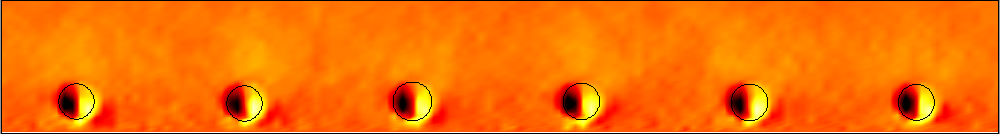}\\ \vspace{0.5mm}
  1.5 D \includegraphics[width=12cm, clip=true]{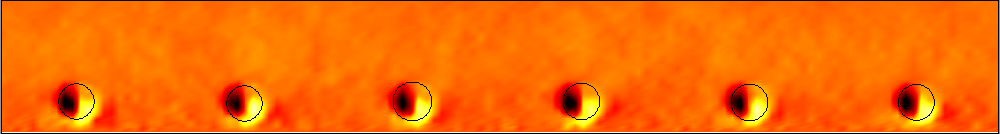}\\ \vspace{0.5mm}
  2.0 D \includegraphics[width=12cm, clip=true]{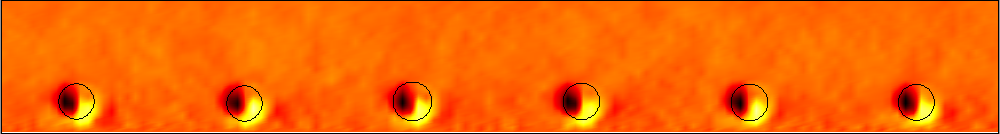}\\ \vspace{0.5mm}
  2.5 D \includegraphics[width=12cm, clip=true]{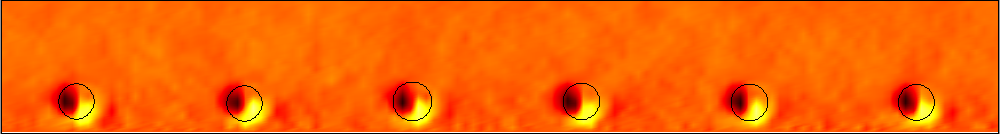}\\ \vspace{0.5mm}
 \hspace{4mm}  \includegraphics[width=9cm, clip=true]{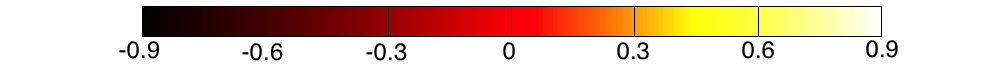}
 \end{center}
\caption{Vertical turbulent humidity flux distribution in $(y,z)$ planes at different streamwise locations, in the wakes of a row of rotors. The distance from the rotor is shown on the left, where $D$ is the rotor diameter.}
\label{f8}
\end{figure}

 The humidity distribution in the vertical direction (i.e., the decrease underneath the hub and increase above the hub) is dictated by the vertical turbulent humidity flux. A parallel can be made with the entrainment of kinetic energy from above layers that was observed in previous studies, except here the humidity is convected in both directions, from and toward the ground. The humidity generated at the ground level (note that we imposed a humidity flux at the bottom boundary) is convected to the upper layers by the rotor wakes, namely by the vertical turbulent humidity flux $\langle w'q' \rangle$. At the same time, the dry air masses (or 'less humid' air) from above the wind farm is convected downward into the wind farm (this is what we could call 'entrainment'). To support this postulation, in figure \ref{f8} we plot contours of the vertical turbulent humidity flux in $(y,z)$ planes at different streamwise locations, starting from 0.1 diameters (in proximity to the rotor) to 2.5 diameters downstream from the rotor disk. It reveals how the humidity is convected from the ground on one side of the rotor, and toward the ground on the other side (the effect of rotation is also captured by these plots). This could be also correlated with the distribution of temperature in a wind farm operating in stable conditions, where it was observed that the rotor wakes slightly heat the air in the region underneath the hub height (while it is known that heating is commonly associated with a decrease in humidity).

\begin{figure}[H]
 \begin{center}
  \includegraphics[width=13cm, clip=true]{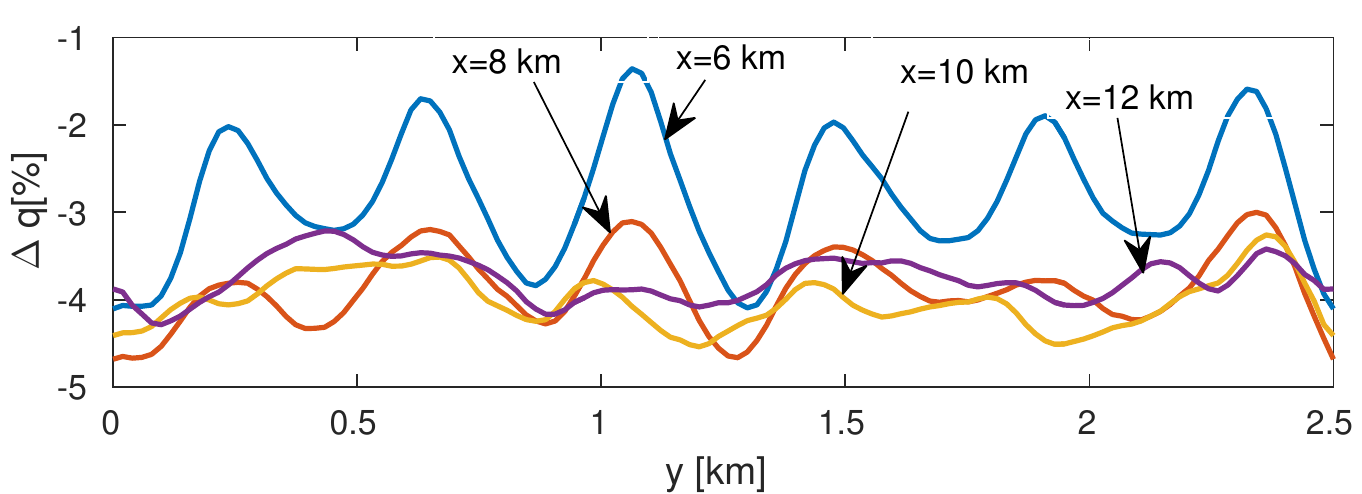} \\
  \includegraphics[width=13cm, clip=true]{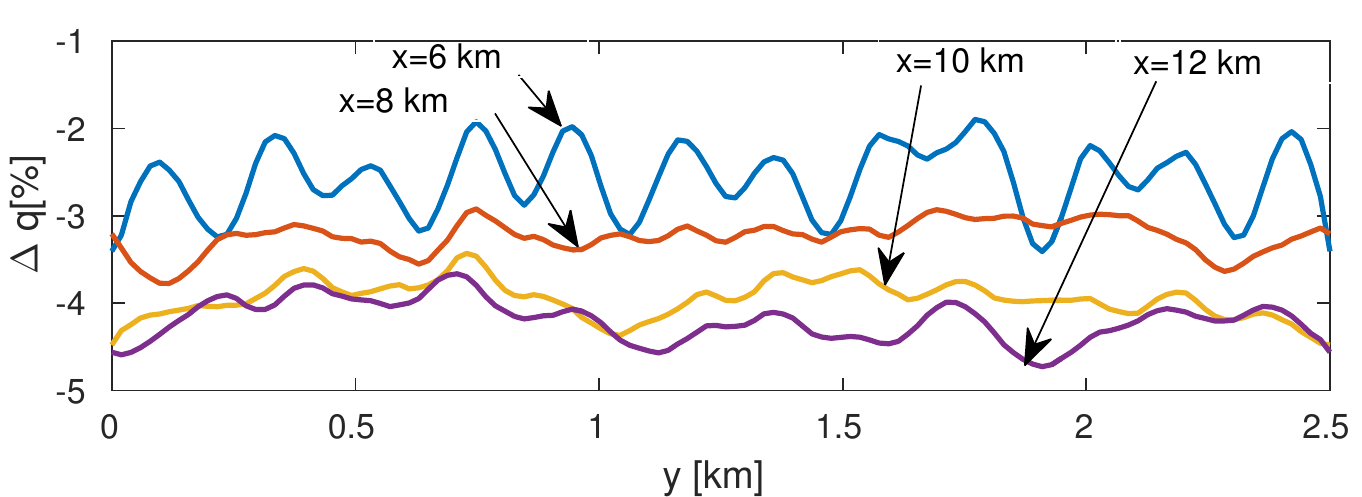}
 \end{center}
\caption{Mean minimum (close to the ground) relative humidity variation in the lateral direction: aligned layout (top); staggered layout (bottom).}
\label{f9}
\end{figure}

\begin{figure}[H]
 \begin{center}
  \includegraphics[width=13cm, clip=true]{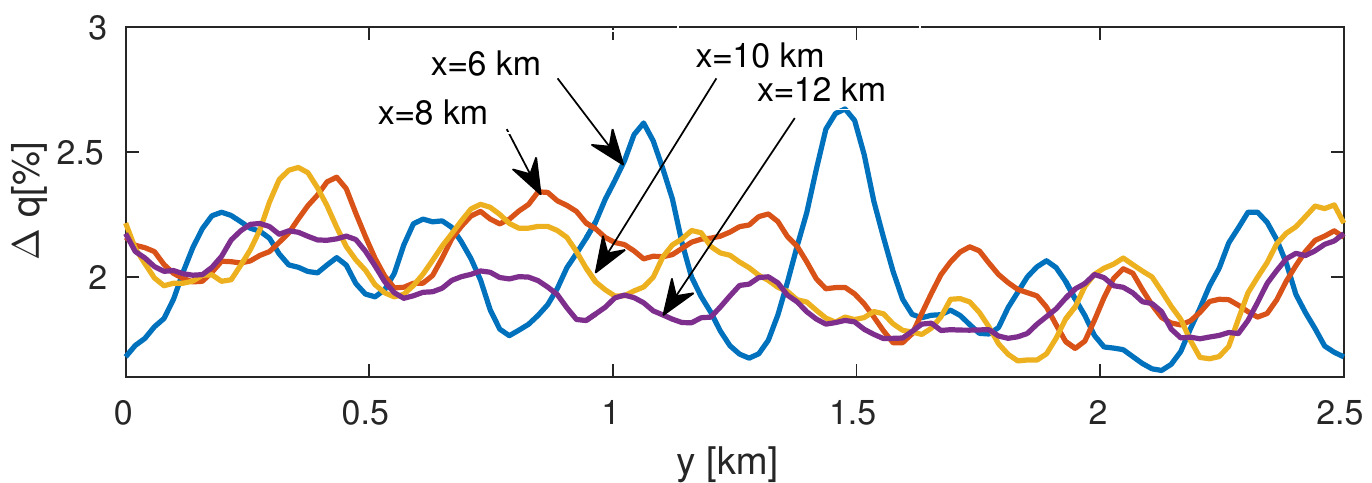} \\
  \includegraphics[width=13cm, clip=true]{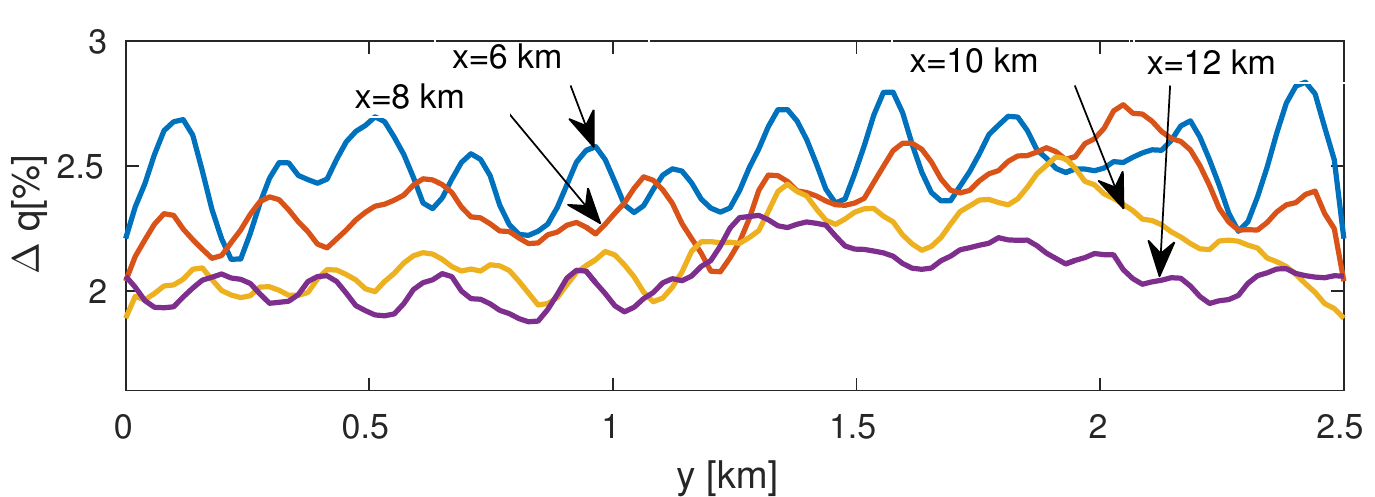}
 \end{center}
\caption{Mean maximum (above the hub) relative humidity variation in the lateral direction: aligned layout (top); staggered layout (bottom).}
\label{f10}
\end{figure}

Finally, in figure \ref{f9} we analyze lateral profiles of the time-averaged relative humidity progression at four selected streamwise locations to determine the impact of the compounded wakes on the humidity distribution in the spanwise direction. For the aligned layout, the effect of the wakes seems to be visible in the profiles that are taken at $x=6$ km and $x=8$ km, but by the time the flow reaches $x=10$ km, the wakes coalesce and thoroughly mix up the humidity. The mixing from the compounding wakes for the staggered farm configuration is even higher, as expected, as seen from the bottom panel of figure \ref{f9}: while the impact of the wakes is predominant at $x=6$ km, by the time the flow reaches $x=8$ km the humidity seems to be well mixed up. The reason for which the wakes' 'footprints' in figure \ref{f9} (the blue curve of the top plot, for example) do not show periodicity is because there are large flow structures in the atmospheric boundary layer that persist for a long time; these flow structures introduce anisotropy in the time averaging taken along practical time intervals (excessively long time intervals should be taken to eliminate the effect of these structures from the data). Similar profiles for the maximum increase of the mean relative humidity progression above the hub, at the same selected streamwise locations, are illustrated in figure \ref{f10}. The above conclusions with respect to the effect of the wakes on the lateral distribution of humidity holds for these profiles as well.

%\subsection{Second level head text}

\section{Conclusions}\label{sec5}

Changes to near-surface relative humidity were analyzed via LES both within and downstream of a large wind farm in aligned and staggered configurations. In keeping with previous observations and simulations made within the array, mixing brought about by turbines change near-surface relative humidity by reducing humidity adjacent to the ground and increasing it aloft.  This investigation shows that this alteration remains well beyond two times the streamwise dimension of the wind farm and, at the surface, is greatest for a staggered layout. Similar to that for TKE, the aligned array configuration allows for a faster recovery of the decrease in relative humidity adjacent to the ground. Along with a faster recovery, the decrease in relative humidity does not extend as far vertically for the aligned configuration but, for both arrangements, extends well above the hub height. While the area of increase in relative humidity aloft rises vertically with downstream distance, the area of greatest increase extends further in the downstream and vertical directions for the aligned array. It was also shown that the humidity distribution in the vertical direction (i.e., the decrease underneath the hub and increase above the hub) is mainly the effect of the vertical turbulent humidity flux, which is positive on one lateral side of the rotor and negative on the other; a consequence of this is a convection of the humidity from and toward the ground, respectively. 

Finally, we advise that any changes of concern with respect to humidity within a wind farm must be considered for long distances downstream of the wind farm as well, although more investigations (either numerically or experimentally) in other different conditions are warranted in this respect.

\end{document}